# Taming intermittent plasticity at small scales

Peng Zhang[a,*], Oguz Umut Salman[b,*], Jin-Yu Zhang[a,*], Gang Liu[a], Jérôme Weiss[c],

Lev Truskinovsky[d], Jun Sun[a]

[a] *State Key Laboratory for Mechanical Behavior of Materials, Xi'an Jiaotong University, Xi'an, 710049, China*
[b] *CNRS, LSPM UPR3407, Université Paris 13, Sorbonne Paris Cité, 93430 Villetaneuse, France*
[c] *IsTerre, CNRS/Université Grenoble Alpes, 38401 Grenoble, France*
[d] *PMMH, CNRS -- UMR 7636, ESPCI ParisTech, 10 Rue Vauquelin, 75005 Paris*

**Abstract** —The extreme miniaturization in modern technology calls for deeper insights into the non-conventional, fluctuation dominated mechanics of materials at micro- to nano-scales. Both experiments and simulations show that sub-micron face-centered-cubic (FCC) crystals exhibit high yield strength, which is, however, compromised by intermittent, power law distributed strain fluctuations, typical of 'wild plasticity'. At macro-scales, the same bulk materials show 'mild plasticity' characterized by bounded, uncorrelated fluctuations. Both anomalous strength and intermittency appear therefore as size effects. While the former is highly desirable, the latter is detrimental because stochastic dislocation avalanches interfere with forming processes and endanger structural stability. In this paper we show that defectiveness, which is used in classical metallurgy to harden materials, can be used at sub-micron scales to suppress intermittent fluctuations. We quantify the coexistence of 'wild' and 'mild' fluctuations in compressed Al alloys micro-pillars, demonstrate that the statistical nature of fluctuations is determined by sample size, and propose quantitative strategies allowing one to temper plastic intermittency by artificially tailored disorder. Our experimental results are rationalized using a theoretical framework which quantifies the competition between external (size related) and internal (disorder related) length scales.

**Key words:** Dislocation dynamics; size effects; plastic fluctuations; hardening

*: These authors contributed equally to this work

Correspondence and requests for materials should be addressed to lgsammer@mail.xjtu.edu.cn (GL) or jerome.weiss@ujf-grenoble.fr (JW) or lev.truskinovsky@espci.fr (LT) or junsun@mail.xjtu.edu.cn (JS)

# 1. Introduction

The classical paradigm of dislocation-mediated plasticity in crystalline solids is that of a smooth *flow* [1, 2] where strain fluctuations are small and uncorrelated. This vision of *mild* plasticity was fundamentally challenged by the discovery that plastic fluctuations may be power law distributed in size and energy [3-6], with clustering in space [7] and time [8]. The fact that dislocations self-organize and plasticity proceeds through collective avalanches implies that the flow is *wild* in the sense of Mandelbrot [9].

The two apparently conflicting pictures of smooth and rough plasticity have been recently reconciled as it was shown that, in bulk materials, mild and wild fluctuations can coexist, with a degree of wildness depending on crystal structure [9]. In hexagonal close-packed (HCP) crystals, long-ranged elastic interactions dominate, leading to cooperative behavior of dislocations. Instead, in face-centered cubic (FCC) crystals, short-range interactions, enhanced by the multiplicity of slip systems, quench plastic avalanches. Plastic flow then proceeds through mainly small and uncorrelated dislocation motions, confined inside the transient microstructural features (e.g. dislocation cells), which give rise to Gaussian (mild) fluctuations. Those coexist with rare power-law distributed (wild) fluctuations associated with brutal rearrangements [9].

In view of the growing interest towards building progressively smaller technological devices, classical approaches of size-independent material engineering have to be reconsidered [1]. In particular, metal plasticity has to be reassessed to meet the demands posed by the manufacturing of components at the micro/nano scales [10] and experiments with ultra-small pillars have become a standard tool in the study of the corresponding fluctuations and size effects [6, 11-14]. Besides the initial observation that "smaller is stronger" [11], it has been recently argued that "smaller is wilder" [9], as, in contrast to Gaussian distributed, small plastic fluctuations in bulk FCC materials [9], scale-free intermittency has been confirmed at micro and nano scales for the same materials by a wealth of experiments [6, 12, 14-16] and simulations [10, 17]. The abrupt strain jumps in quasi-statically loaded micro-/nano-components endanger structural stability and the associated unpredictability raises serious challenges for plastic-forming processes [17]. It has been realized that tempering plastic deformation at ultra-small scales requires new approaches going beyond the phenomenological continuum theory [18].

In bulk materials, grain boundaries (GBs) hinder the propagation of dislocation avalanches, introducing grain-size related upper cut-offs on their size distribution [19]. At micro- and nano-scales, the level of poly-crystallinity cannot be controlled with the same confidence as in bulk materials [20], which limits our ability to use GB for mitigating size-induced intermittency. Considering these limitations, we focus on a different strategy, motivated by recent simulation which showed that quenched disorder may suppress scale-free behavior in bulk materials [21]. We use the fact that the pinning strength of solutes and precipitates can be artificially tailored within metals by simple aging treatments [22]. At sub-micro scales, quenched disorder gives the potential of smoothing plastic flow [23-26]. However, this effect has not been quantified so far in terms of avalanche statistics.

To explore this strategy, we systematically study the effects of miniaturization on strain fluctuations in Al-alloys single crystals strengthened by different types of solutes or precipitates in the conditions when the grain size is not a relevant length scale of the problem. We experimentally *quantify* the "smaller is wilder" effect in pure crystals, reporting an evolution from mild plastic behavior at large pillar diameters $L$, to wild plasticity at small $L$. We then provide evidence that the transition between mild to wild regimes shifts towards smaller $L$ with the increase of the pinning strength of quenched disorder. Translating the pinning strength into a characteristic length scale $l$, we show that the competition between *external* (due to $L$) and *internal* (due to $l$) scale effects can be quantified by a single nondimensional parameter $R = L/l$ allowing one to collapse the data for materials with different degree of defectiveness on a single curve. We rationalize this collapse within a simple theoretical framework that builds an unexpected bridge between wildness and strength. Our study suggests specific quantitative strategies for controlling intermittency in sub-μm plasticity.

## 2. Experimental procedures

### *2.1 Materials*

The experiments were performed on four different types of Al crystals: (i) pure Al, (ii) Al-0.3wt.%Sc alloy with Sc solute clusters (referred to as Al-Sc cluster in Fig. 1a), (iii) Al-0.3wt.%Sc alloy with fine sphere-like Al$_3$Sc precipitates of size ~ 3-8 nm (referred to as Al-Sc precipitate in Fig. 1b), and (iv) Al-2.5wt.%Cu-0.1wt.%Sn with coarse plate-like $\theta'$-Al$_2$Cu precipitates of diameter ~ 10-40 nm (referred to as Al-Cu-Sn in Fig. 1c). The pure Al, Al-0.3wt.%Sc alloy, and Al-2.5wt.%Cu-0.1wt.%Sn alloys were respectively melted and cast in a stream argon, by using 99.99 wt.% pure Al, mast Al-50 wt.% Cu alloy, 99.99 wt.% pure Sn, and mast Al-2.0 wt.% Sc alloy. The cast Al-Sc ingots were solutionized at 921 K for 3 h and then quenched in cold water. Immediately after quenching, one part of the Al-Sc ingots was aged at relatively low temperature of 523 K for 8 h to form Sc clusters. The other part of the Al-Sc ingots was aged at high temperature of 623 K for duration of 24 h, in order to precipitate $Al_3Sc$ particles. The cast Al-Cu-Sn ingots were solutionized at 823 K for 3 h, followed by a cold water quench and subsequently aged at 473 K for 8 h to precipitate plate-like $\theta'$-$Al_2Cu$ particles. Minor addition of micro-alloying element Sn was used to catalyze the precipitation of $\theta'$-$Al_2Cu$ particles with relatively uniform size and homogeneous distribution.

## 2.2 Microstructure characterization

Three-dimensional atom probe tomography (3DAP) analyses were performed using an Imago Scientific Instruments 3000HR local electrode atom probe (LEAP) to examine the three-dimensional distribution of Sc atoms in the Al-Sc cluster alloy (Fig. 1a). The 3DAP experiments routine can be found elsewhere [27]. The reconstruction and quantitative analysis of 3DAP data were performed using the IVAS 3.4.3 software. The precipitates in aged alloys were quantitatively characterized by using transmission electron microscope (TEM) and high resolution TEM (HRTEM) (Fig. 1). TEM foils were prepared following standard electro-polishing techniques for Al alloys. Quantitative measures of number density and size of the precipitates are reported as average values over more than 200 measurements. The foil thickness in the beam direction was determined through convergent beam electron diffraction pattern [28]. Volume fraction of the particles was determined by using corrected projection method [29]. Details about the microstructural measurements can be found in our previous publications [2, 22].

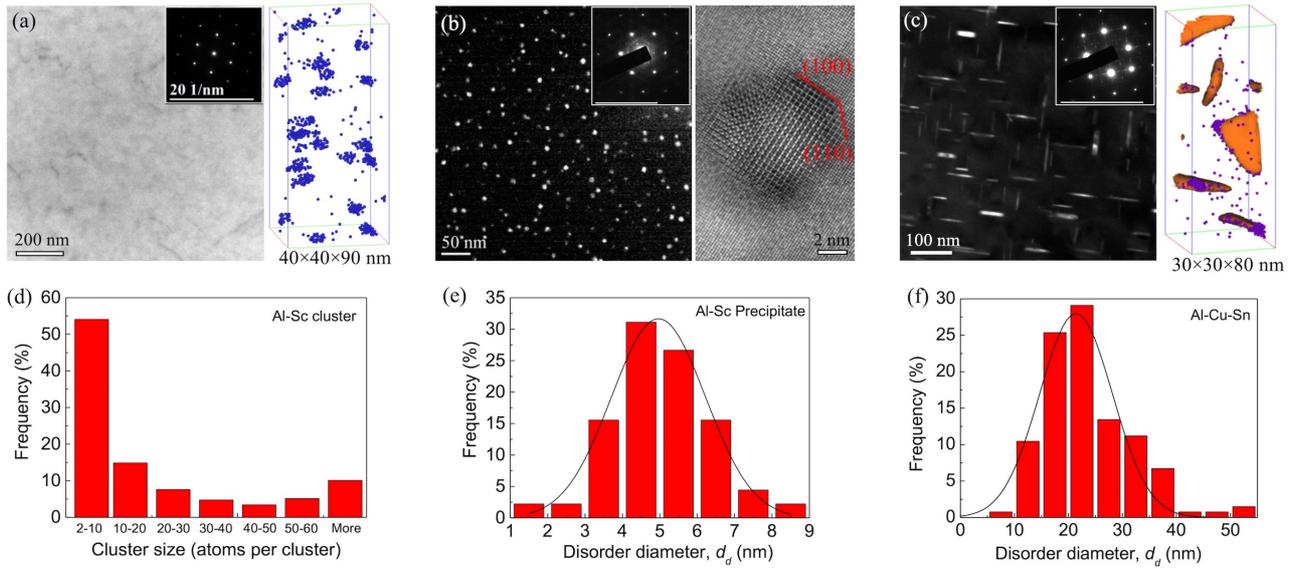

**Figure 1. Microstructural characteristics of Al alloys. (a)** Bright field TEM image (left) and 3DAP result (right) for the Al-Sc alloy. No precipitates can be observed but a large number of Sc atom clusters are detectable (referred as Al-Sc cluster alloy). The blue points represent Sc atoms. **(b)** Dark field TEM image (left) and HRTEM image (right) for the Al-Sc alloy dispersed with spherical $Al_3Sc$ precipitates with average diameter ~5 nm (referred as Al-Sc precipitate alloy). **(c)** Dark field TEM (left) and 3DAP (right) images for the Al-Cu-Sn alloy with plate-like $\theta'$-$Al_2Cu$ precipitate with a diameter ~25 nm (Al-Cu-Sn). The purple points represent Sn atoms and orange ones are $\theta'$ precipitates. Insets in the TEM images are the corresponding selected area diffraction patterns. The statistical results of obstacle size in Al-Sc cluster (a), Al-Sc precipitate (b), and Al-Cu-Sn alloys (c) are shown in **(d), (e), (f)** respectively.

## 2.3 Micro-pillar fabrication and compression

Micro-pillars with the same <110> orientation were fabricated by focus iron beam (FIB) within <110>-oriented grains, which were locked by Electron back scattered diffraction (EBSD) on the electro-polished surface of each material (Fig. 2). The micro-pillar diameter ranged from about 500 up to about 6000 nm, and the height-to-diameter ratio of all the micro-pillars was kept between 2.5:1 and 3.5:1 (Fig. 2c). Fabrication procedures for micro-pillars have been detailed in previous publications [11, 13].

The micro-compression tests were performed on a Hysitron Ti 950 with a 10 μm side-flat quadrilateral cross-section diamond indenter. The micro-pillars were compressed under the displacement-controlled mode at a strain rate of $2\times10^{-4}$ $s^{-1}$ up to 20% strain. The cross-sectional area at half height of the pillar and the initial height were used to calculate true stresses and strains, following a well-known methodology proposed in Ref. [13].

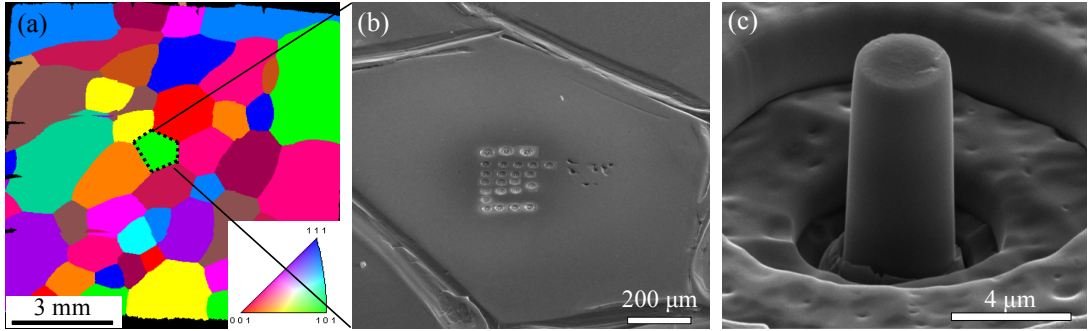

**Figure 2. Fabrication methodology of micro-pillar. (a)** Representative EBSD orientation map of Al-Cu-Sn alloy, from which the <110>-orientated grain, marked by dash lines along the grain boundaries, is locked to fabricate micro-pillar inside. **(b)** A series of micropillars are fabricated within the <110>-orientated grain marked in (a). **(c)** Magnified SEM image to show the morphology of micro-pillar.

## 3. Characterization of tested materials

### *3.1 Microstructure of aged Al alloys*

Fig. 1a, b and c show representative microstructural images of Al-Sc cluster, Al-Sc precipitate, and Al-Cu-Sn alloys, respectively. In the Al-Sc cluster alloy, careful examinations found no perceptible precipitates formed (TEM image of Fig. 1a). According to previous works [30], the aging temperature should be greater than ~ 623 K in order to precipitate $Al_3Sc$ second phase particles. In the present work, the Al-Sc cluster alloy was aged at a much lower temperature (523 K). However, abundant Sc solute clusters can be detected from 3DAP analyses, as shown in Fig. 1a. Core-Linkage (CL) algorithm [27] was used to quantify the size of Sc clusters. The statistical results given in Fig. 1d show that most of the detected Sc clusters contain less than 10 Sc atoms. As the dislocations can shear through the solute clusters, the interaction between the solute clusters and dislocations is relatively weak.

In the Al-Sc precipitate alloy that was aged at 623 K for 24 hours, a large number of nano-sized $Al_3Sc$ particles are precipitated and dispersed in the matrix, see the TEM image of Fig. 1b. The coherent $Al_3Sc$ precipitates have an equilibrium shape of Great Rhombicuboctahedron [30], with a total of 26 facets on the {100}, {110} and {111} planes (refer to the HRTEM image of Fig. 1b). Considering these precipitates as spherical particles, their diameter distributions were quantitatively measured. The statistical results are presented in Fig. 1e, with an average diameter of about 5.0 nm. Previous work on $Al_3Sc$ particle strengthening [31] has shown that the transition from dislocation shearing of precipitates to Orowan's looping occurs for a critical precipitate diameter of about 4.2 nm. Hence, in the present case, the interactions between dislocations and the Al$_3$Sc precipitates, dominated by bypassing mechanisms, are strong.

The Al-Cu-Sn alloy was aged at 473 K for 8 h, which led to the precipitation of plate-like $\theta'$-Al$_2$Cu particles (see the TEM image of Fig. 1c). The micro-alloying element Sn was doped to refine the distribution of $\theta'$ precipitates by promoting $\theta'$ nucleation [29]. Representative 3DAP results shown on Fig. 1c illustrate the micro-alloying mechanism: fine Sn particles are firstly formed by Sn atoms segregation, and then these Sn particles provide preferential nucleation sites for the $\theta'$ precipitates. Due to the Sn micro-alloying effect, the $\theta'$ precipitates in the Al-Cu-Sn alloy have much reduced sizes compared with those in the Al-Cu counterpart [32]. In addition, the Sn-promoted $\theta'$ precipitates also have a narrower distribution in size. This optimization of the size of precipitates improves the repeatability of micro-pillar testing, and therefore is suitable for investigating the precipitate-dislocation interactions at small length scales. The statistical results in Fig. 1f show an average diameter of ~25 nm for the $\theta'$ precipitates. Since the $\theta'$ precipitates are intrinsically shear-resistant [29, 32], the $\theta'$ precipitates exhibit a typical bypassing strengthening mechanism with a strong precipitate-dislocation interaction.

The measured parameters of the precipitates/solute clusters, including sizes, density, volume fraction, have been summarized in Table 1.

### *3.2 Evaluation of obstacle resistance to dislocation motion*

We will show that obstacle resistance to dislocations motion, which arises from disorders (solution atoms, clusters and precipitates), forest dislocations, and lattice friction for the studied materials, is an important controlling parameter in the following analysis. This pinning strength to dislocation motion can be evaluated from experimental measurements of the yield strength at 0.2% offset under tension in bulk polycrystalline samples. We used bulk sample testing rather than micro-pillar testing to determine pinning strengths in order to eliminate the inevitable external size effect on strength that exists in case of micro-pillars. The bulk samples were cut from as-cast ingots that were exposed only to heat treatments and underwent no warm/cold deformation. This resulted in: (a) nearly equiaxial grains, free of texture, ensuring a relatively homogeneous and isotropic deformation, and (b) a large average grain size ($L_g \sim$ 1 mm, Fig. 2a), leading to a small grain boundary strengthening. For coarse-grained materials, we can evaluate the pinning strength ($\tau_{pin}$) as follows:

$$\tau_{pin} \approx \sigma_y/M - kL_g^{-1/2} \tag{1}$$

where $\sigma_y$ is the bulk yield strength derived from the stress-strain curves under uniaxial tension (right half of Fig. 3a), $M$ the Taylor factor (3.06 for FCC metals [33]), and $k$ the Hall-Petch constant for Al alloy (60 MPa μm$^{1/2}$ [34]). The evaluated values are listed in Table 1. Note that these values are effective strengths, including the contributions from disorder strengthening, forest dislocation strengthening, and lattice friction.

To support these experimental estimates, we also calculated the pinning strength based on dislocation strengthening theory and microstructural statistics. The lattice friction stress $\tau_l$ is ~1.4 MPa [35]; the forest dislocation strengthening $\tau_\rho \sim \alpha Gb\sqrt{\rho_f}$ is around 3.45MPa (taking an initial forest dislocation density of ~10$^{12}$ m$^{-2}$ for the well annealed crystals), where $G$ is the shear modulus of Al (24.7 GPa [32]), $b$ the Burgers vector (0.286 nm) and $\alpha \approx 0.5$. Thus, the pinning strength in pure Al is ~ 4.85 MPa. For Al alloys, the main contribution comes from disorders. In case of non-shearable precipitates, the pinning strength is [22, 29]:

$$\tau_p = \frac{Gb}{2\pi\sqrt{1-\upsilon}} \frac{1}{\lambda} \ln\left(\frac{\pi d_d}{4b}\right) \text{ (for sphere-like precipitates)} \tag{2a}$$

$$\tau_p = \frac{Gb}{2\pi\sqrt{1-\upsilon}} \frac{1}{\lambda} \ln\left(\frac{0.981\sqrt{d_d t_d}}{b}\right) \text{ (for plate-like precipitates)} \tag{2b}$$

, where $\upsilon$ is the Poisson's ratio (0.33 [32]), and $\lambda$ the interparticle spacing. The value of $\lambda$ can be obtained by using the experimentally measured precipitate parameters in Table. 1 [29]:

$$\lambda = \frac{1.075}{\sqrt{N_v d_d}} - \frac{\pi d_d}{4} \text{ (for sphere-like precipitates)} \tag{3a}$$

$$\lambda = \frac{1.2669}{\sqrt{N_v d_d}} - \frac{\pi d_d}{8} - 1.061 t_d \text{ (for plate-like precipitates)} \tag{3b}$$

where $N_v$ is the number density, $d_d$ is the precipitates' diameter, and $t_d$ the thickness of plate-like precipitates. Note that, for all alloys, the average interparticle spacing is significantly smaller than the micro-pillar diameters (from 500 nm to 3500 nm) (Table 1). This ensures that atom clusters and precipitates dispersed within the micro-pillars can effectively pin the dislocations. The microscopic estimates of pinning strength $\tau_{pin} = \tau_p + \tau_l + \tau_\rho$ for Al-Sc precipitate (38.05 MPa) and Al-Cu-Sn (70.25 MPa) are in excellent agreement with the estimates based on experimental measurements mentioned above.

**Table 1 | Statistical results on microstructure features and evaluation of internal scale.**

| Materials | Disorder Diameter $d_d$ (nm) | Disorder Thickness $t_d$ (nm) | Number density $N_v$ ($10^{22}$ m$^{-3}$) | Volume fraction % | Disorder spacing $\lambda$ (nm) | Shearing or bypassing | Pinning strength (MPa) | internal scale $l$ (nm) |
|---|---|---|---|---|---|---|---|---|
| Pure Al | - | - | - | - | -- | -- | 4.5(0.7) | 1536.9 |
| Al-Sc cluster | 1.81(0.04)+ | - | 131.00(5.01) | 0.41(0.04) | 20.6(0.6) | Shearing | 27.6(2.2) | 250.6(18.5) |
| Al-Sc precipitate | 5.08(0.24) | - | 1.86(0.11) | 0.13(0.03) | 107.0(6.0) | Bypassing | 38.3(1.3) | 180.6(5.9) |
| Al-Cu-Sn | 24.51(3.02) | 2.31(0.21) | 0.99(0.05) | 1.21(0.42) | 66.4(6.1) | Bypassing | 70.5(3.5) | 98.1(4.6) |

+: Individual cluster diameters were calculated as the Guinier's diameter ($d_d = l_d\sqrt{5/3}$), where $l_d$ is the diameter of gyration[27].
The values in the bracket stand for the measurement error of TEM examination (microstructural parameter) and bulk tension tests (pinning strength).

## 4. Statistics of intermittent plasticity

### *4.1 Deformation behaviors of micro-pillars*

The ordering of obstacle pinning strengths (pure Al (~ 4.5 MPa) < Al-Sc cluster (~ 28 MPa) < Al-Sc precipitate (~38 MPa) < Al-Cu-Sn(~ 70 MPa) dictates the hierarchy of flow stresses for both bulk materials and micro-pillars (Fig. 3a). The stress-strain curves for pure Al micro-pillars, showing almost no bulk-like strain hardening, appear jerky over the pillar diameter range analyzed in our experiments. A representative deformation curve for pure Al micro-pillar with 2000 nm diameter is presented in Fig. 3a, illustrating the frequent occurrence of strain bursts. In contrast, Al-Cu-Sn crystals deform (flow) smoothly and strain-harden within the same diameter range, as shown on Fig. 3a. The behavior of Al-Sc cluster and Al-Sc precipitate micro-pillars lay in between these two end-member scenarios, with a jerkiness clearly decreasing with increasing pillar diameter (typically shown in Fig. 3b).

Moving beyond the "jerky" or "smooth" terminologies generally used in the literature to describe these stress-strain curves, we propose below an objective quantification of the difference between these two behaviors. While clearly different from the highly discontinuous deformation where plasticity reveals itself through abrupt strain bursts (Fig. 3c), smoother deformation curves also reveal some strain bursts scattered among extended segments of seemingly continuous deformation (Fig. 3d). Although the stress may slowly fluctuate during the smooth segments of the stress-strain curves, plastic bursts cannot be unambiguously identified in the corresponding strain rate-time signals (bottom half of Fig. 3d), implying a much lower dissipation rate compared with the episodes containing strain avalanche. These two kinds of plastic processes, being fundamentally different in terms of dissipation rate, will be interpreted as wild and mild plasticity [9]. Our observations not only confirm the existence of these two limiting behaviors, i.e., wild and mild plasticity, but also suggest the possibility of their coexistence [9]. The key to understand the tradeoff between wild and mild deformation is therefore to quantify their relative importance.

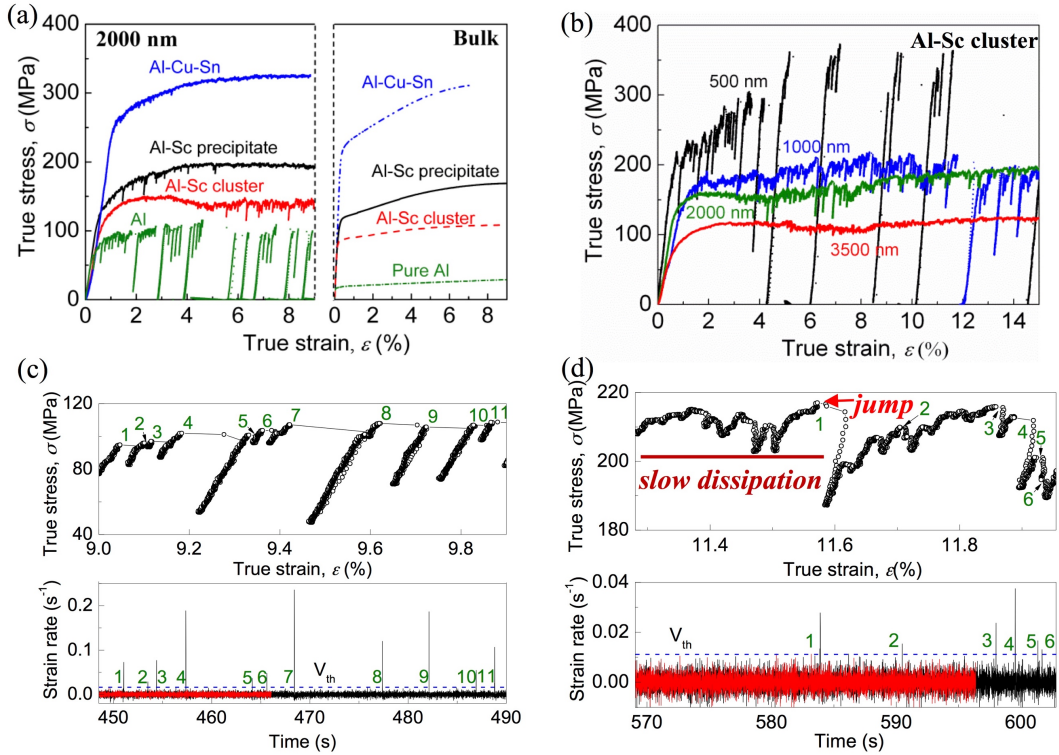

**Figure 3. Representative stress-strain curves. (a)** Typical stress-strain curves for pure Al and Al alloys. (Left) for 2000 nm diameter micro-pillars, we show the effect of alloying/disorder on deformation. Stronger disorder suppresses the jerky character of deformation. (Right) for bulk samples, we show the effect of disorder on yield stress and hardening. **(b)** Typical stress-strain curves of the Al-Sc cluster micro-pillars with different sample sizes demonstrating the effect of the external scale. A segment of stress-strain curves (upper half) and the corresponding strain rate-time signal (black line, bottom half) for a 2000 nm diameter pure Al **(c)** and Al-Sc cluster **(d)** micropillars. The strain rate-time signal during initial elastic loading is also given for comparison (red line); the strain rate target ($2 \times 10^{-4}$ s$^{-1}$) has been removed, therefore the red line represents a background noise. The strain rate thresholds $V_{th}$ are used to separate the bursts (marked by 1, 2…) from background noise. In the pure Al sample, plasticity is exclusively released through abrupt strain bursts. In contrast, a significant part of plastic deformation is released in a smooth manner without any detectable strain burst ("slow dissipation") for the Al-Sc cluster sample. Some plastic bursts are also visible ("jumps").

### *4.2 Quantification of the external size and disorder effects on intermittent plasticity*

We explicitly define an "avalanche" as a plastic process characterized by a dissipation rate much greater than the imposed loading rate. Based on this definition and in view of the control mode provided by the loading device [36] (Hysitron Ti 950, see details in Supplement Materials (SM)), a dislocation avalanche manifests itself by both (i) a displacement/strain burst away from the strain-rate target (Fig. S1a, b and Fig. S2a in SM) and (ii) an abrupt stress drop (Fig. S1a, c and Fig. S2c in SM). These two different manifestations can be separately used to distinguish wild from mild fluctuations through an objective methodology detailed in SM. The cumulative effect of extracted wild fluctuations, normalized by the total imposed deformation, will be used as the wildness parameter $W$. The results obtained via displacement-rate signals (Fig. S2a and b) and force-rate signals (Fig. S2c and d) are in close agreement (compare Fig. 4 with Fig. S3 in SM), arguing for the robustness of our methodology. We focus on the role of external size and defectiveness on plastic fluctuations, comparing samples of different size and/or different materials, and all the results presented below (wildness $W$, distributions of strain bursts,..) correspond to signals integrated over the entire stress-strain curves. Some characteristics, such as the forest dislocation density, may be expected to evolve with deformation, whereas a possible evolution of the distribution of strain avalanches with increasing strain has been discussed recently in Refs. [4, 6, 37]. We do not address these points here, which are left for future work.

Fig. 4a summarizes two main results of our work. First, we experimentally *quantify* the "smaller is wilder" phenomenon, which we interpret as an *external* size effect. Second, we show that the crossover range between wild (large $W$) and mild (small $W$) plasticity can be shifted (or even suppressed down to sub-µm scales, see Al-Cu-Sn) by introducing high pinning-strength disorder. Such "dirtier is milder" phenomenon reveals the presence of a disorder-related characteristic scale and shows that the corresponding *internal* size effect competes with the more conventional *external* size effect. Recall that in bulk materials, a stronger degree of wildness correlates with a smaller value of a non-universal exponent $\kappa$ in the power law distribution of strain bursts $P(s) \sim s^{-\kappa}$ [9]. The maximum likelihood analysis [38] of plastic avalanches in micro-pillars confirms this trend down to sub-µm scales (Fig. 5) and reveals a universal (material independent) relation between $W$ and $\kappa$ (Fig. 4b). This crossover behavior illustrates the gradual transition between Gaussian (mild) and power-law (wild) fluctuation regimes, and provides a unifying perspective on micro-plasticity. The limiting behaviors are observed in Al-Cu-Sn alloy, where the distribution of the (few) detected strain bursts is almost Gaussian, and in pure Al (≤3500 nm), showing almost purely power-law distributed bursts with exponent $\kappa$ approaching (for the smallest sizes) the previously reported value [6, 17, 39, 40] close to the one predicted by the mean field theory, $\kappa=1.5$ [3].

Note that this transition from power-law to Gaussian distributions should not be confused with an upper cut-off of a power law scaling. Indeed, we could expect a "trivial" finite-size effect to impact our displacement burst distributions from the upper side [4, 17]. However, in this case, this cut-off should be more pronounced upon decreasing the sample size, in opposition with what is shown on Fig. 5. In addition, comparing the likelihoods between power-law and power-law with upper cut-off distributions ($P(s) \sim s^{-\kappa}$ and $P(s) \sim s^{-\kappa} exp\left(-\frac{s}{s_{max}}\right)$, where $s_{max}$ is the upper cut-off amplitude) by the likelihood ratio test [38], possible upper cut-offs were hardly detectable. This is likely due to the limited statistics of our datasets, but also reinforces the reliability of our estimation of the exponent $\kappa$.

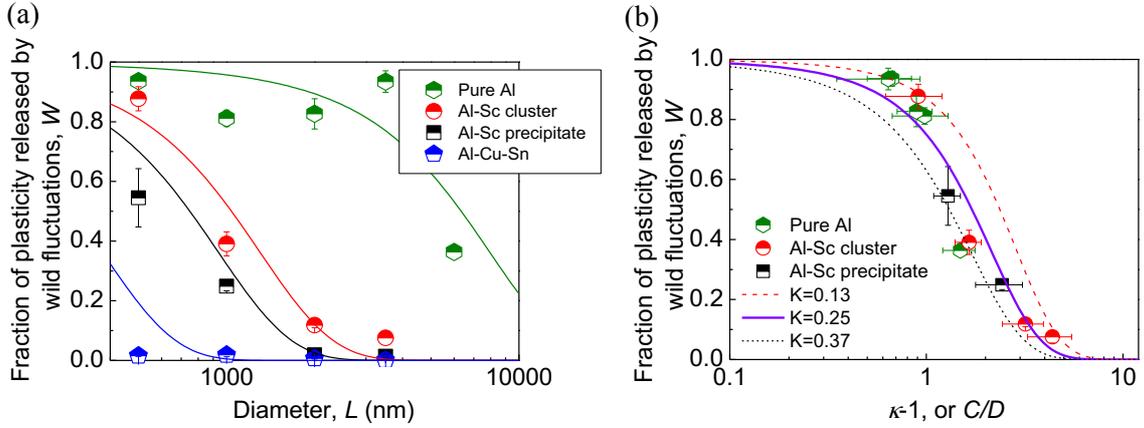

**Figure 4. The dependence of the fraction of plasticity released by wild fluctuations, $W$, on the nature of the disorder, micro-pillar diameter (a) and power law exponent $\kappa - 1 = C/D$ (b). (a)** Fraction of plasticity released by wild fluctuations, $W$, as a function of sample diameter ($L$) for the pure Al and Al alloys micro-pillars. "Smaller is wilder" and "dirtier is milder" phenomena can be clearly observed. The plotted $W(L)$ solid lines are calculated from our model, using quantitative estimates for the parameters $C$ and $D$ (see section 6.2.2). Error bars are derived from measurements on multiple samples. **(b)** Universal (material independent) relationship between $W$ and the power law exponent $\kappa - 1 = C/D$ (see text for details). Error bars represent the uncertainty of power law exponents fitted by the maximum likelihood analysis [38].

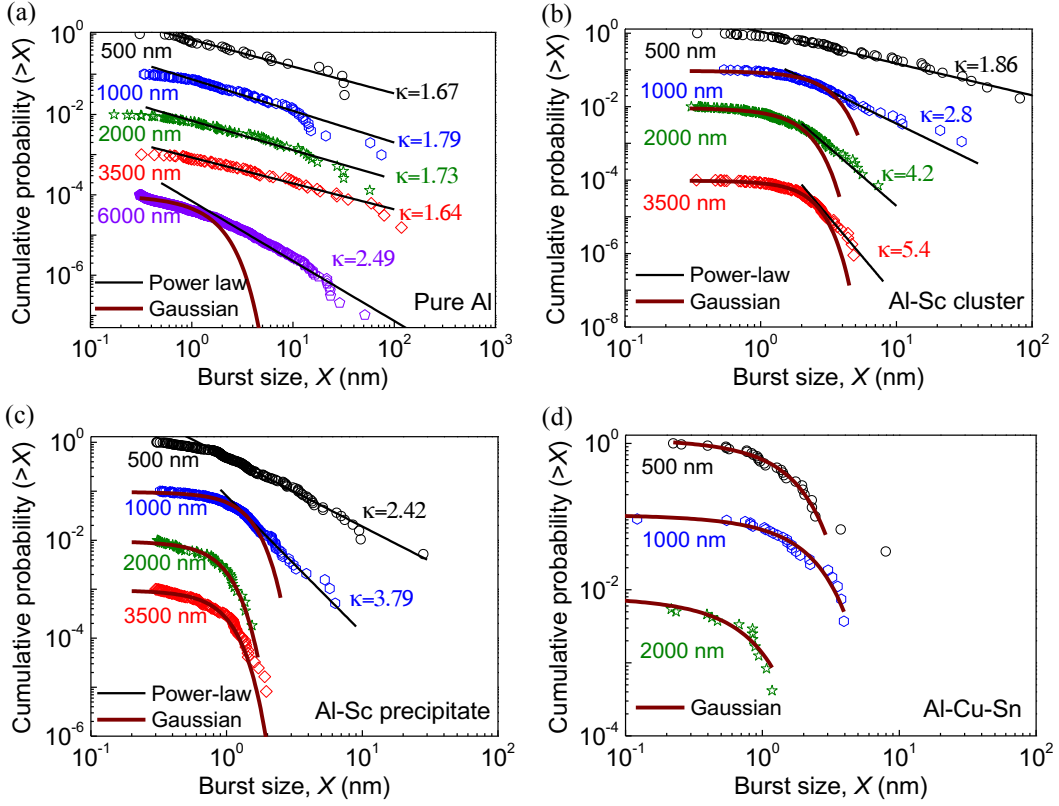

**Figure 5. Distributions of detected displacement burst sizes in pure Al (a), Al-Sc cluster (b), Al-Sc precipitate(c), and Al-Cu-Sn micro-pillars (d)**. In (b) and (c), the burst sizes follow a Gaussian distribution at small sizes, and a power law distribution at large sizes. The power law exponent is a function of external size and disorder. A crossover from a power law to a Gaussian distribution can be observed as we increase the diameter and/or the pinning strength of the disorder.

## *4.3 Nondimensional parameter R*

The "dirtier is milder" phenomenon quantified above implies the existence of a disorder-related characteristic scale that competes with the *external* size effect. To interpret our experiments, we define this internal length scale as $l = Gb/\tau_{pin}$ for pure metals and alloys, where $\tau_{pin}$ is the pinning strength of all kinds of obstacles (see section 3.2). It can be perceived as a consequence of a very simple dimensional analysis, i.e., $l/b = G/\tau_{pin}$. In a more physical interpretation, $l = Gb/\tau_{pin}$ describes the length scale below which the dislocation-dislocation elastic interaction stress (scaling as $Gb/l$) equals the dislocation-obstacle interaction stress $\tau_{pin}$ [41]. Hence $l$ indeed controls the transition from "endogeneous" (dislocation-dislocation) to "exogeneous" (dislocation-obstacle) interaction. Using this definition of internal length scale, our experimental results suggest that the crossover form mild to wild plasticity is governed by a single nondimensional parameter $R = L/l$ (Fig. 6a).

The scaling collapses of our data (Fig. 6a and c) show that indeed $W = W(R)$ and $\kappa = \kappa(R)$, therefore the decrease of $L$ can be compensated by the proportional increase of $\tau_{pin}$. From this interpretation, we can construct a new regime map (Fig. 7) to display the coupled effect of external size and disorder on intermittency for different types of defectiveness. In this regime map, the fluctuations are characterized by two important parameters, i.e., power-law tail exponent $\kappa$ and wildness $W$, which are directly linked with each other through the universal relation observed on Fig. 4b.

It should be mentioned that we do not have grains in our samples, so we essentially assume that the corresponding characteristic scale $L_g$ is larger than any of the above scales. Another important scale is, of course, $l_m = \rho_m^{-1/2}$, where $\rho_m$ is the representative mobile dislocation density solving our stochastic differential equation in the model section below. This so defined dislocation spacing cannot be controlled and is changing in the process of deformation. We implicitly assume that in our wild regime $l_m \sim L \leq l \leq L_g$ (dislocations do not see the defects, while interacting with the surface: external size is in control) and in our mild regime $l_m \sim l \leq L \leq L_g$ (dislocation interaction with defects is essential while the surface plays a secondary role: internal size is in control). The crossover from mild to wild regime takes place when internal and external scales become of the same order: $L \sim l$.

For pure metals with negligible lattice friction, the tunable obstacles are represented by immobile forest dislocations of density $\rho_f$, hence in this case $\tau_{pin} \sim Gb\sqrt{\rho_f}$ and $R \sim L/l_f = L\sqrt{\rho_f}$, where $l_f = 1/\sqrt{\rho_f}$ is proportional to the effective mean free path of mobile dislocations [42, 43]. Therefore, on the one hand, our parameter $R \sim L\sqrt{\rho_f}$ is fully compatible with scaling properties of dislocation systems in pure metals deduced from similitude principles [44]. On the other hand, the same parameter $L/l_f$ has also been shown to control the nature of hardening at small scales for pure metals [39, 45] (see detailed discussions in the next section), implying a possible connection between the average hardening behavior and statistical fluctuations. In the following, we will precise the nature of this correlation and reveal that the single parameter $R$ can unify both the hardening and wildness map, for both pure metals and alloys.

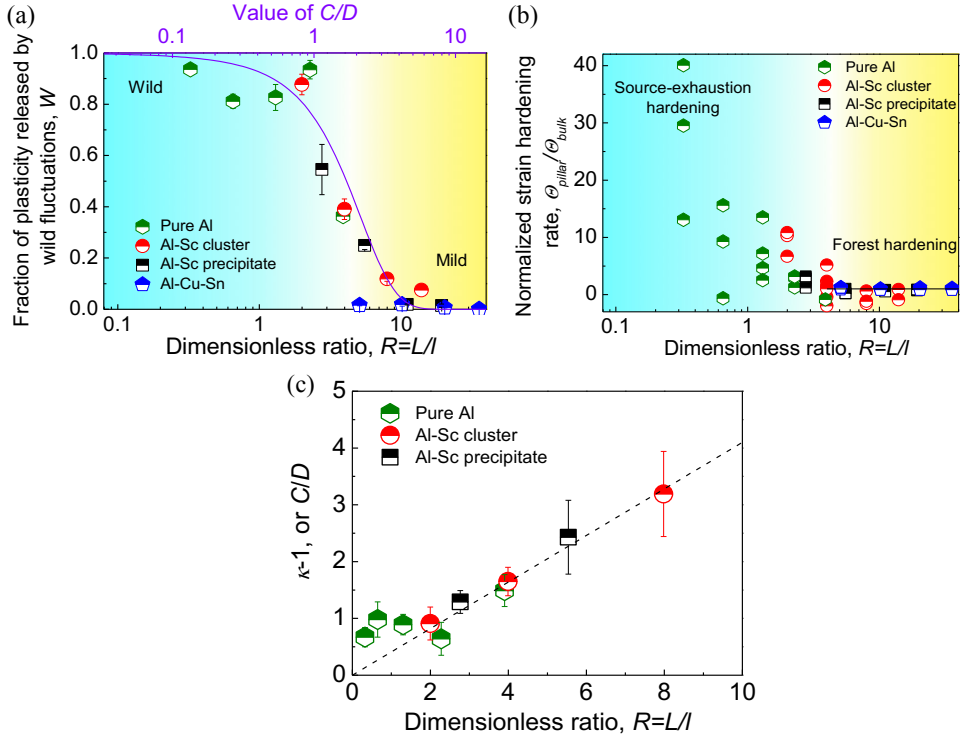

**Figure 6. The fraction of plasticity released by wild fluctuations, $W$, the normalized strain hardening rate by the bulk counterpart, $\Theta_{pillar}/\Theta_{bulk}$, and the power law exponent, $\kappa - 1 = C/D$, as a function of dimensionless ratio, $R$. (a)** All data of Fig. 4a collapse into a master curve. The solid curves in (a) are calculated from the model, using values of parameter $K = 0.25$. **(b)** The hardening transition occurs around $R \approx 5$, regardless of special materials and in excellent agreement with the wild-to-mild transition in (a). **(c)** A linear relationship between $R$ and $\kappa - 1$ or $C/D$ is observed.

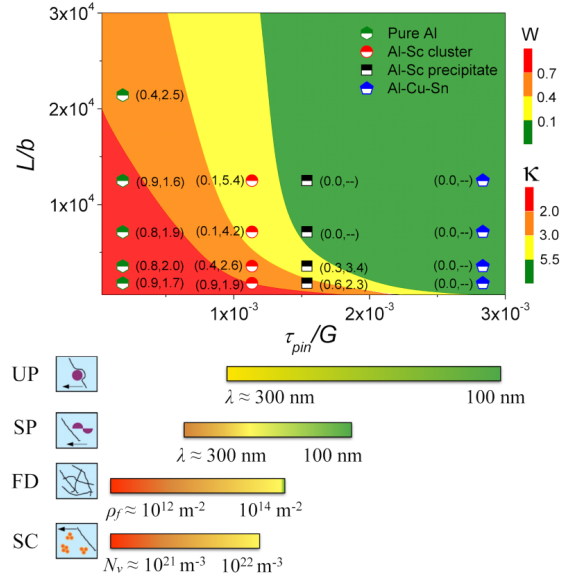

**Figure 7. Tuning map showing the $W$ and $\kappa$ contours as a function of external size $L/b$ and pinning strength $\tau_{pin}/G$.** Corresponding to the pinning strength, a defectiveness spectrum is developed by unifying the unshearable particles (UP, like the $Al_2Cu$ and $Al_3Sc$ precipitates in present work), shearable particles (SP, like fine and coherent $Al_3Sc$ precipitates), forest dislocations (FD, like in the present pure Al), and solute clusters (SC, like the Sc solute cluster in present work; $N_v$ is the cluster number density). The numbers inside the brackets represent $W$ and $\kappa$ respectively. The color bar below illustrate the range of $W$ and $\kappa$ at $L/b = 2\times10^4$.

## 5. Correlations between the average behaviors and the statistics of fluctuations

*5.1 The link between forest hardening and mildness*

In order to examine a possible link between the strain hardening behavior and the statistics of fluctuations, the quantification of strain hardening rate (SHR) and its evolution with internal/external length scale should be firstly addressed. As micro-pillars generally display jerky deformations and serrated strain-stress curves, the evaluation of SHR from traditional methods used in bulk sample testing is difficult. Therefore, following previous works [46-48], we estimated the SHR by $\Theta_{pillar} = \frac{\sigma_{5\%}-\sigma_{2\%}}{5\%-2\%}S^2$, where $\sigma_{5\%}$ and $\sigma_{2\%}$ are the stresses corresponding to the 5% and 2% strains in the strain-stress curves, respectively, and $S$ is the Schmid factor (0.408 for <110> orientation). Strain-stress curves of bulk polycrystalline samples under tension were used to estimate bulk SHR for comparison with micro-pillars. The resolved stress $\tau = \sigma/M$ and resolved strain $\gamma = \varepsilon M$ at slip system are used to calculate the SHR following the same evaluation method as in micro-pillars by $\Theta_{bulk} = \frac{\tau_{\gamma_5}-\tau_{\gamma_2}}{\gamma_5-\gamma_2}$, where $M$ is the Taylor factor, $\gamma_5 = \frac{5\%}{S}$ and $\gamma_2 = \frac{2\%}{S}$.

The dependence of SHR as a function of pillar diameter $L$ is shown on Fig. 8, for the different alloys. For pure Al, the SHR value of the pillar of diameter $L = $ 6000 nm is close to the bulk value, whereas values obtained for smaller sizes are extremely scattered, another illustration of the jerkiness of plastic flow [49]. The Al-Sc cluster micro-pillars with $L < $ 2000 nm and the Al-Sc precipitate micro-pillars with $L < $ 1000 nm show scattered SHR values as well, with a mean value greater than that of their bulk counterpart. This suggests a breakdown of size-independent forest hardening mechanism in small samples, in relation with weak mutual dislocation reactions and dislocation storage in small sample [45, 50], and the activation of dislocation sources dominating the plasticity [45, 46, 51]. To sustain flow in this case where weak dislocation sources are exhausted, much higher stress is required to activate the remaining truncated ones [49, 51, 52]. Moreover, the stochastic distribution of initial dislocation sources in the small volume leads to a large scatter in the SHR at small $L$, whereas scattering decreases when SHR values reach bulk values towards large $L$. Such a transition from classical forest hardening to source-exhaustion hardening upon decreasing the size $L$ has been recently studied in experiments, by mean-field modeling [39], and by DDD simulations [45]. Our results shown on Fig. 8 are fully consistent with these studies.

However, beyond the effect of external size $L$ on the hardening mechanisms, we show here that strong disorder can also shift this transition towards smaller $L$, as it does for the wild-to-mild transition (see Fig. 4a). Indeed, we show in Fig. 8 that the hardening transition scale, larger than 3500 nm for pure Al, decreases from about 2000 nm for Al-Sc cluster micro-pillars to about 1000 nm for the Al-Sc precipitate samples, and less than 500 nm for Al-Cu-Sn. The hierarchy observed for the mildness transition is recovered and, although a precise estimation of the hardening transition scale is difficult from these results, the values given above are in good agreement with the length scales at which most of the wild fluctuations are suppressed (see Fig. 4a). In other words, a material starts to strain-harden when it becomes mild. This correspondence is further supported by the scaling collapse of Fig. 6a and b, showing that our non-dimensional parameter $R$ controls both the wild-to-mild transition and the nature of hardening.

The results shown in Fig. 6 suggest that these two transitions are the manifestations of the same underlying phenomenon. In case of pure single crystal, Alcalá et al. [39] proposed that the non-dimensional ratio $L/l_{eff}$ controls the transition from forest hardening to source-exhaustion hardening, where $l_{eff}$ is an "effective" mean free path for dislocations. These authors estimated $l_{eff}$ from an expression accounting for different types of dislocation interactions. In their case, given that one can neglect the effect of lattice friction on $\tau_{pin}$, the mean free path for mobile dislocations $l_{eff}$, which scales as $1/\sqrt{\rho_f}$ [42, 43], can be identified with our internal length scale $l$ (see section 4.3). Consequently, for pure metals, the ratio $L/l_{eff}$, which controls the nature of hardening [39], can be identified with our parameter $R = L/l$. This parameter, in turn, accounts for the wild-mild transition (Fig. 6a), arguing in favor of a non-incidental overlap of the wild-to-mild transition and the transition between two different hardening mechanisms.

Our Fig. 6 extends this correlation beyond pure metals, and suggests that alloying increases dislocation storage capacity, favoring the formation of dislocation entanglements and enhancing forest hardening. Instead, it disfavors dislocation avalanches and wild plastic fluctuations. In summary, according to our unified scenario, external and internal size effects control, through the non-dimensional ratio $R = L/l$, the nature of both the *average* behavior (hardening) and the statistics of *fluctuation*s.

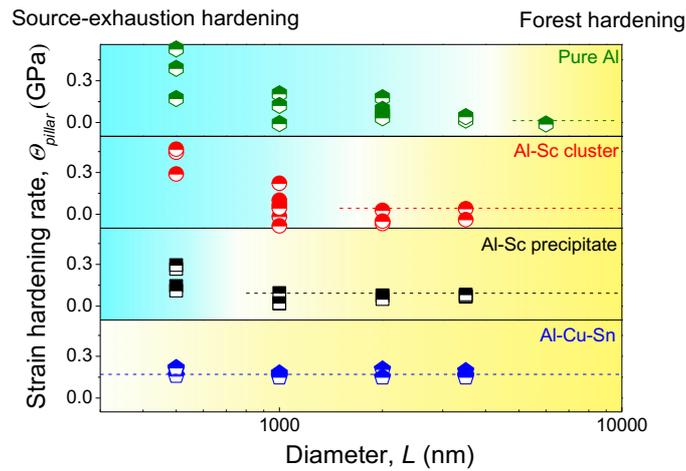

**Figure 8. Strain hardening rate for micro-pillars.** The small samples with weak obstacles display more likely source-exhaustion hardening, characterized by $L$-dependent SHRs in high value and large scattering. Introducing obstacles suppresses source-exhaustion hardening, and extends forest hardening to small diameter. The dotted lines represent the SHR of bulk samples for comparison. The critical diameter for hardening transition is in good agreement with the wild-to-mild transition diameter (as shown in Fig. 4a).

## *5.2 The link between deformation morphology and mildness*

The SEM images of Fig. 9 show the deformation morphologies of micro-pillars after uniaxial compression tests. Small micro-pillars (diameter of 1000 nm) are compared with larger ones (diameter of 3500 nm), to demonstrate the effect of external sample size. Pure Al, Al-Sc cluster, Al-Sc precipitate, and Al-Cu-Sn micro-pillars are compared to illustrate the influence of disorder characteristics. The results show that: (i) in the 1000 nm micro-pillars, plastic deformation is predominantly accommodated by localized deformation on one or two slip plane(s), except for Al-Cu-Sn in Fig. 9g, indicating that plastic deformation in these small-sized micro-pillars exhibiting a large wildness (see Fig. 4a) is strongly anisotropic. This anisotropy of plastic flow for the symmetrical multiple slip system is actually a breakdown of Schmid's law [53] . (ii) On the reverse, in 3500 nm-sized Al-Cu-Sn micro-pillars, more uniform plastic deformation is found, with barrel-like deformation morphology. The large size allows significant dislocation motion before annihilation at free surface [52, 54] and so dislocation sources to be activated within the bulk on multiple slip planes [46]. 3D Dislocation networks are formed, leading to the uniform plastic deformation associated with mild fluctuations and radial dilation under compression (see Fig. 9h). (iii) In the pure Al micro-pillar with a large diameter of 3500 nm (Fig. 6b), plastic deformation is still concentrated on a single slip plane, as, in the absence of disorder, dislocations can easily escape to be annihilated at free surfaces.

In order to quantify the strain homogeneity as a function of external size and disorder, the expanding ratio, defined as the percentage change of diameter after 20% deformation, is used. The diameter measurements were always performed at the half height of the pillars, before deformation and after 20% deformation respectively. All the measurements on diameter were performed along the same <001> orientation to avoid the possible influence of expanding anisotropy. A clear trend is shown in Fig. 10, that is, larger is the diameter or stronger is the disorder, more homogenous is the deformation, consistent with the direct SEM observations.

Further exploring the correlation between expanding ratio and mildness, it is strikingly found that all data roughly collapse into a master curve, regardless of the materials and external size (Fig. 10b). This indicates a same origin for strain homogeneity and mildness, linked to the hindering of avalanches and the prevalence of short-range interactions.

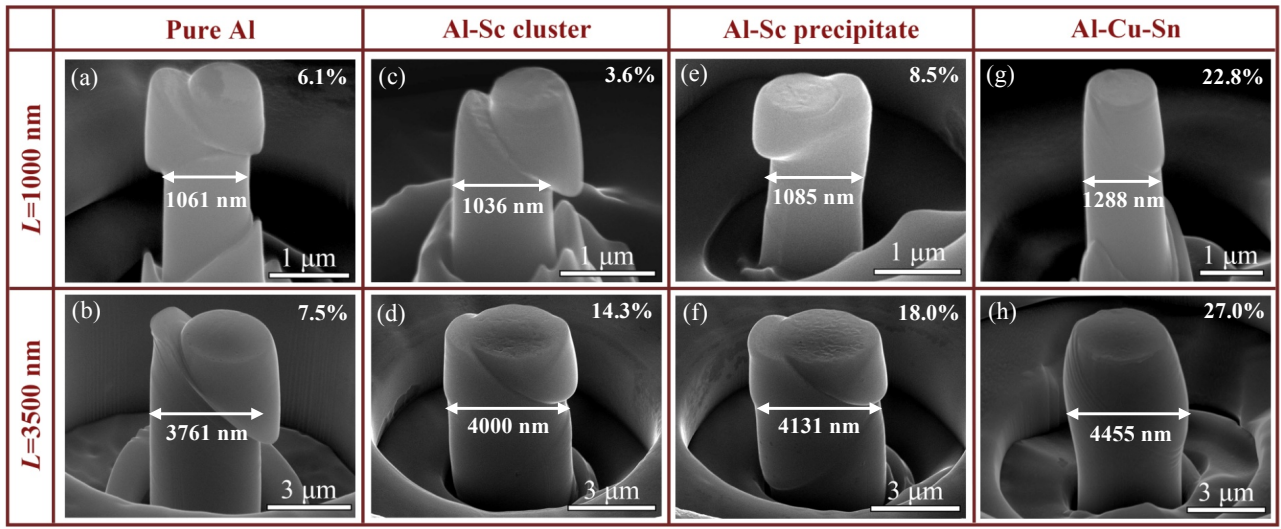

**Figure 9. Effects of external size and internal disorder on deformation morphology.** SEM images showing the deformation morphology of pure Al (**(a)** and **(b)**), Al-Sc cluster (**(c)** and **(d)**), Al-Sc precipitate (**(e)** and **(f)**), and Al-Cu-Sn (**(g)** and **(h)**) micro-pillars with diameter of 1000 nm (**(a)**, **(c)**, **(e)**, and **(g)**) and of 3500 nm (**(b)**, **(d)**, **(f)**, and **(h)**). The diameter expanding ratio is given at the top right corner.

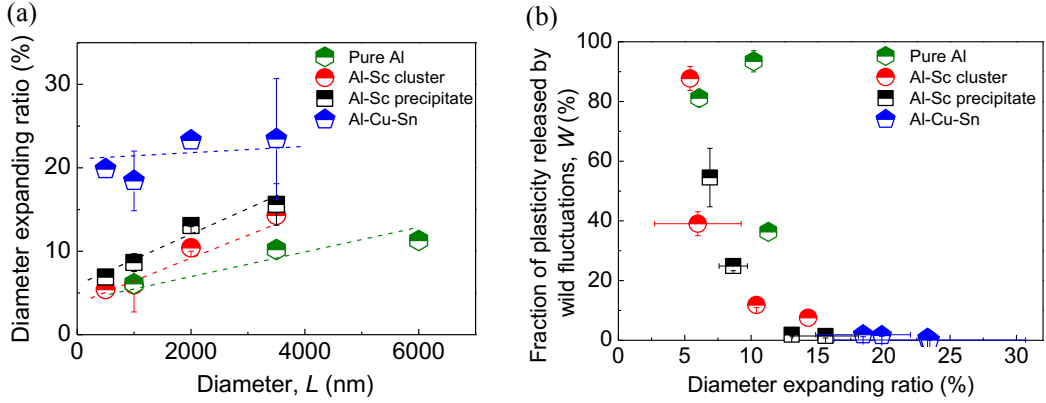

**Figure 10. Diameter expanding ratios for micro-pillars and their correlation with *W*. (a)** For small sized pillars with weak disorder, where the fluctuations dominate the deformation (Fig. 3b), the strain is highly localized in few slip planes, without apparent swelling, i.e. the diameter expanding ratio is smaller, and vice versa. **(b)** Relationship between the degree of wildness *W* and the diameter expanding ratio. All data roughly collapse into a master curve, regardless of materials and external size, arguing for a common origin.

*5.3 Brief summary*

The correlations revealed above uncover a complex dependence of dislocation dynamics on sample size, disorder and crystal symmetry. Wild plasticity is rooted in anisotropy (Fig. 10b): small-sized pure crystals exhibit single slip flow (Fig. 9) even when multiple slip planes are active in bulk samples [46]. It can be also related to the smaller role played by short-range interactions among dislocations due to practical absence of locks, junctions, etc. In particular, the probability of mutual dislocation reactions, resulting in entanglements and immobilization, diminishes with the sample size [13], and the size-induced high yield stress can compromise their stability [50]. This prevents the formation of a usual self-induced microstructure, which suppresses intermittency in high symmetry bulk materials [9]. In this way one can rationalize the observation that "smaller is wilder" and justify the observed correlation of isotropic deformation with mildness.

The intriguing link between the mild-to-wild transition and the transition from the regime of forest hardening to the regime of source-exhaustion hardening (Fig. 6a and b) indicates that a material becomes mild when forest hardening is favored. Our analysis of disordered samples supports these findings, suggesting that the high-strength defects can mimic forest dislocations by quenching avalanches [21] while simultaneously strengthening the material (Fig. 3a). The disorder impedes dislocations from reaching free surface sinks and facilitates entanglements, making plasticity more isotropic (i.e. multi-slip, Fig. 9) and promoting forest hardening (Fig. 8). The data collapse suggests that both hardening-related and fluctuations-related transitions are controlled by the competition between the external scale $L$ and the internal scale $l$ which can be quantified in terms of a single nondimensional parameter $R = L/l$ (Fig. 6).

The correlations between strain hardening mechanisms, deformation morphology and fluctuation behavior are rooted in a common origin of all these effects which can be quantified in terms of the probability of mutual dislocation reactions and the feasibility of the formation of dislocation entanglements.

## 6. Model

### *6.1 Model structure and prediction of the universal relation W-κ*

Our experimental results find a simple interpretation in the framework of a mean field model for the density of mobile dislocations $\rho_m$, where the dynamics under *constant* stress is described by the stochastic equation [9]:

$$\frac{d\rho_m}{d\gamma} = A - C\rho_m + \sqrt{2D}\, \rho_m \xi(\gamma) \tag{4}$$

Here the time-like parameter $\gamma$ is chosen to coincide with the applied shear strain, $A > 0$ is the net nucleation rate taking into account a negative contribution from surface annihilation, $C > 0$ is the rate of mutual annihilation/immobilization of the dislocation pairs, $D$ is the intensity of self-induced fluctuations and $\xi(\gamma)$ is a standard white noise with $\langle\xi(\gamma)\rangle = 0$, $\langle\xi(\gamma_1)\xi(\gamma_2)\rangle = \delta(\gamma_1-\gamma_2)$. The heuristic model (4), where we focus on fluctuations at a fixed stress and do not address strain hardening, can be viewed as a spatially resolved mesoscopic closure of the continuum mechanical model of plasticity. We assume that the local yielding thresholds depend on a single dislocation density and long range stochastic interactions are implemented in the form of a multiplicative mechanical noise. Similar approaches, relying on the idea of 'mechanical temperature' as a description of the fluctuating local stress, have been used in the modeling of athermal amorphous plasticity [55] and in the representations of crystal plasticity as a noise induced transition [56]. Various ways of representing correlations in the mechanical action of the noise in such models are expected to capture the persistent nonlocal rearrangements and avalanche-like processes triggered by local instabilities [57]. In crystal plasticity such stochastic constitutive relations still await to be validated by the rigorous upscaling procedures based on microscopic models [57, 58]. To link our model with mechanical measurements, we can use Orowan's relation $d\gamma = \rho_m b v dt$. Recalling that we consider a fixed external stress condition, the average dislocation velocity $v$, set by the applied stress [59], can be taken as a constant. Thus, the distribution of strain fluctuations will be exactly reflected by the fluctuations of mobile dislocation density $\rho_m$ solved in our model.

The stationary probability distribution predicted by Eq. (4) takes the form $P(\rho_m) = \frac{\left(\frac{A}{D}\right)^{C/D} e^{-\frac{A}{D\rho_m}} \rho_m^{-\left(1+\frac{C}{D}\right)}}{\Gamma\left(\frac{C}{D},0\right)}$, where $\Gamma(z,x) = \int_x^\infty t^{z-1} e^{-t} dt$. We can then write $\kappa - 1 = C/D$ and define the degree of wildness as $W = \int_{\rho_{min}}^{\infty} P(\rho_m) d\rho_m$ where $\rho_{min}$ is the threshold distinguishing between wild and mild fluctuations. It is natural to specify $\rho_{min}$ by the condition $\exp\left(-\frac{A}{D\rho_{min}}\right) = K$ where $K = 1$ corresponds to a pure power law. At fixed $K$ we obtain an important relation between the degree of wildness $W$ and the power law exponent $\kappa - 1 = C/D$ of the long tail describing intermittent fluctuations of the dislocation density

$$W = 1 - \frac{\Gamma\left(\frac{C}{D}, \log\left(\frac{1}{K}\right)\right)}{\Gamma\left(\frac{C}{D}, 0\right)} \qquad (5)$$

The fact that the ratio $A/D$ drops from Eq. (5) is in agreement with the experimental scaling collapse, indicating that the degree of wildness $W$ depends only on the exponent $\kappa$ (see Fig. 4b). Quite remarkably, the predicted curves $W(C/D)$ have the same sigmoidal shape as experimental data, and by choosing the value $K = 0.25$, we achieve an almost perfect fit; the $K$ dependence in the range of interest is relatively weak (Fig. 4b).

### *6.2 The external size and disorder effect in the model*

So far, an analytical study of Eq. (4) allowed to give specific predictions in terms of scaling exponent $\kappa$ and degree of wildness $W$. In this section, we will relate our model parameters to $L$ and $l$ in order to rationalize the main results from our experiments, i.e., the sample size and disorder effect on $W$ and $\kappa$.

**6.2.1 The interpretation of the model parameters**

The *model* has two characteristic densities $\rho_C = A/C$ and $\rho_D = A/D$, which gives rise to two characteristic lengths $l_C = \frac{1}{\sqrt{\rho_C}} = \sqrt{C/A}$ and $l_D = \frac{1}{\sqrt{\rho_D}} = \sqrt{D/A}$. The ratio of these two length scales, $r = \frac{l_C}{l_D} = \sqrt{C/D}$, is the main dimensionless parameter of the model. Our *analytical* study shows that the fraction of strain released in wild fluctuations $W$ depends only on $r$, in other words that $W = W(C/D)$. On the other hand, our *experimental* study shows that $W = W(R)$ where the parameter $R = L/l$ is also dimensionless. By comparing the functions $W(r)$ and $W(R)$ we find that $R \sim r^2$ which means that $\kappa - 1 = \frac{C}{D} \sim \frac{L}{l} = R$. This linear relation has been confirmed experimentally, as shown in Fig. 6c.

Our parameter $C$ characterizes the rate of *mutual* annihilation/immobilization, which Gilman calls 'stalemating' [60]. It is fundamentally different from surface annihilation of individual dislocations, which negatively contributes to our parameter $A$ ($\frac{d\rho_m}{d\gamma} = -\frac{1}{bL}$) [61]. In fact, as we have shown, $A$ does not affect either the degree of wildness $W$ or the avalanche distribution. In small samples, the immobile configurations, such as dislocation dipoles and locks, are hard to form (or easy to break) under the external size induced high stresses, meaning that an extrinsic size effect applies on parameter $C$. Available 3-D discrete dislocation simulations (Fig. 8 in Ref. [50]) suggests an approximately linear relation between the "dislocation reaction (number) per volume" and the external size $L$. Combined with a similar dislocation density during straining for all $L$ (Fig. 5 in Ref. [50]), this implies that the "dislocation reaction rate" increases linearly with $L$, that is $C \sim L/b$. In the next section, we will justify this scaling based on a detailed microscopic approach and estimate the value of the proportionality coefficient.

The connection between $D$ and classical (non stochastic) models of dislocation dynamics is, by nature, less straightforward. In view of the interpretation of the parameter $C$, i.e., $C \sim L/b$, we can conclude that in the range where $C/D \sim L/l$ (Fig. 6c), the mechanical temperature of the system [62] $D \sim l/b = G/\tau_{pin}$ diminishes with increasing pinning strength of obstacles and increases with stiffness responsible for the long range interactions.

The proposed interpretation of the model parameters is consistent with the effect of crystal structure on wildness in bulk materials [9]. For instance, in HCP materials like ice with essentially single slip plasticity, forest hardening is absent. Therefore $C$ accounts for mutual annihilation only and remain small even at macro scale. Combined with a very small lattice friction, this also implies a small pinning strength and thus $D$ is large independently of $L$. Hence, bulk ice crystals remain wild, and most probaly outside the diagram shown in Fig. 6c as the correlation $C/D \sim L/l$ should be valid only over a limited range of $L/l$. At sufficiently small $L/l$ (large degree of wildness), our exponents for the aggregate distribution (integrated over a range of stress values) appear to be saturated around $C/D$=0.5 (Fig. 6c). This value is close to the prediction of the mean field theory [3, 6] but may also be an indicator of a smaller stress-integrated exponent obtained in recent DDD modeling [4]. At large values of $L/l$ (small degree of wildness), $C$ should saturate towards a bulk value $C_{bulk}$; the associated departure from the scaling $C/D \sim L/l$ is hardly detectable in our experimental data, as detected avalanches become too rare to allow an estimation of $\kappa$.

### 6.2.2 The physical expression of the model parameters, and the link with the size effect on strength

To justify the scaling $C \sim L/b$ and estimate the value of $C$, we start by writing the expressions for the rates of mutual annihilation, dipole formation, or lock formation proposed by Roters et al. [63]:

$$\dot{\rho}_m^-(annihil) = 2\chi_{annihil} \frac{\dot{\gamma}}{b} \frac{1}{n} \rho_m \tag{6a}$$

$$\dot{\rho}_m^-(dipol) = 2(\chi_{dipol} - \chi_{annihil}) \frac{\dot{\gamma}}{b} \frac{1}{n} \rho_m \tag{6b}$$

$$\dot{\rho}_m^-(lock) = 4\chi_{lock} \frac{\dot{\gamma}}{b} \frac{n-1}{n} \rho_m \tag{6c}$$

Here $\dot{\gamma}$ is the shear strain rate, $\chi_{annihil}$ is a critical distance below which two dislocations with antiparallel burgers vectors can annihilate, and $\chi_{dipol}$ and $\chi_{lock}$ are critical distances for the spontaneous formation of dipoles and locks/junctions, respectively. Annihilation can take place with the help of cross-slip and/or dislocation climb. As our tests were performed at room temperature, cross-slip is likely the dominant controlling mechanism in Eq. (6a). $n$ is the number of active slip systems, under the assumption of an equal density of moving dislocations on each of these systems. The combination of terms due to annihilation Eq. (6a) and dipole formation Eq. (6b) yields $\dot{\rho}_m^-(annihil + dipol) = 2\chi_{dipol} \frac{\dot{\gamma}}{b} \frac{1}{n} \rho_m$. As $\frac{\dot{\rho}_m^-}{\dot{\gamma}} = \frac{d\rho_m^-}{d\gamma}$, we get:

$$\frac{d\rho_m^-}{d\gamma}(annihil + dipol) = \frac{2\chi_{dipol}}{b} \frac{1}{n} \rho_m = C_{annihil+dipol} \rho_m \tag{7a}$$

$$\frac{d\rho_m^-}{d\gamma}(lock) = \frac{4\chi_{lock}}{b}\frac{n-1}{n}\rho_m = C_{lock}\,\rho_m \tag{7b}$$

By accounting for the elastic interactions between dislocations, or by invoking line tension calculations, the critical distances $\chi_{dipol}$ and $\chi_{lock}$ can be translated into critical effective shear stresses acting on dislocations, below which dipoles or locks are stable [64]:

$$\tau_{eff}(dipol) = \frac{Gb}{8\pi(1-v)\chi_{dipol}} \tag{8a}$$

$$\tau_{eff}(lock) = \frac{Gb}{2\pi\chi_{lock}} \tag{8b}$$

Combining Eq. (7) and (8), we obtain for the mutual annihilation/immobilization rate $C$:

$$C = \frac{G}{\pi}\left(2 - \frac{2}{n} + \frac{1}{4(1-v)n}\right)\frac{1}{\tau_{eff}} \tag{9}$$

We write the effective shear stress allowing to unlock an immobile configuration as $\tau_{eff} = \tau_{yield} - \tau_{pin}$, where $\tau_{yield}$ is the yield strength of micro-pillars estimated at 0.2% of plastic strain in our compression experiments. We note that when $n$ is large, the contribution of lock/junction formation in Eq. (9), $(2 - \frac{2}{n})$, strongly dominates the contribution due to annihilation and dipole formation, $(\frac{1}{4(1-v)n})$.

At this stage, a connection between the size effect on wildness, and the well-documented size effect on strength can be made. Although various power law exponents have been proposed to describe the scaling relation between $\tau_{yield}$ and $L$ for different materials [52, 65], our experimental data justify a material-independent linear relation $\tau_{yield} - \tau_{pin} \sim Gb/L$ (Fig. 11a), which is consistent with the re-evaluation of a large set of published experimental data [66]. Actually, the data collapse shown in Fig. 11a implicitly supports the source truncation mechanism [46, 67]. In this case, the effective stress $\tau_{yield} - \tau_{pin}$ required to activate a dislocation source of length $\lambda_s$ goes as $1/\lambda_s$ [67], whereas recent DDD simulations [68, 69] argue for a dislocation source length scaling as external size, $\lambda_s \sim L$. The combination of these relations yields the observed size effect in Fig. 11a. The scattering of source lengths contributes to the scatter of measured effective stress [67], but doesn't change the overall scaling trend. Combining Eq. (9) and our observed correlation between $\tau_{yield} - \tau_{pin}$ and $1/L$ (Fig. 11a), we verify the scaling $C \sim L/b$.

From the knowledge of the proportionality coefficient between $\tau_{yield} - \tau_{pin}$ and $1/L$, equal to 40.8 Pa.m, we obtain the proportionality coefficient in the relation $C \sim L/b$ once $n$ is known. Our micropillars were compressed along the <110> direction, so there are two slip planes and four equal slip systems ((111) <10-1>, (111) <01-1>, (11-1) <101> and (11-1) <011>) with the same Schmidt factor. For pure Al, Al-Sc cluster and Al-Sc precipitates, concentrated slip bands are generally observed (Fig. 9), suggesting $n \approx 2$. Instead, the much more homogeneous transversal deformation of Al-Cu-Sn samples suggests that $n = 4$ is more reasonable in this case. The proportionality coefficient in the relation $C \sim L/b$ changes slightly from 0.065 for $n = 2$ to 0.087 for $n = 4$. Following these numbers, we obtain for $C$ the values ranging from ~100 for pure Al 500 nm-micropillars, to ~1000 for Al-Cu-Sn 3500 nm-micropillars. The observed correlation $\kappa - 1 = C/D \sim L/l$, with a dimensionless proportionality coefficient equal to 0.41, yields $D \sim l/b$ with a proportionality coefficient equal to 0.15 for $n = 2$. Our estimate shows that $D$ ranges from ~100 for Al-Sc precipitate to ~800 for pure Al.

The obtained numerical values of the parameter $C$ and $D$ bring the model in surprisingly good agreement with experimental data (Fig. 4a). However, the $C/D$ value might saturate towards the mean-field value of 0.5 for very small $R$ (Fig. 6c). In this case, the above calculations would slightly underestimate $C/D$ in this range. We also note that the crystal orientation effect on wildness can be reflected in the parameter *n*. In case of single-slip orientation where $n \approx 1$, the formation of locks/junctions is almost negligible, and the calculated coefficient of $C \sim L/b$ is only 0.020, much smaller than in our multi-slip orientation case (0.065 to 0.087). This move the wild-to-mild transition to larger sizes, in agreement with micro-compression experiments of Ni crystals with <-269> orientation showing power-law distributed plastic bursts [12].

Finally, and most importantly, the observed correlation between the power law exponent $\kappa - 1 = C/D$ and the ratio $\tau_{eff}/\tau_{pin}$ shown on Fig. 11b confirms the established relation $C \sim \frac{L}{b} \sim \frac{G}{\tau_{yield} - \tau_{pin}}$ and $D \sim \frac{l}{b} = \frac{G}{\tau_{pin}}$, and builds a conceptual bridge between the size effects on strength in one hand, and wildness on the other hand. This scaling $\frac{\tau_{eff}}{\tau_{pin}} \sim \frac{D}{C} \sim \frac{1}{R}$, which shows that our parameter $R$ controls the size effects on wildness (Fig. 6a), hardening mechanism (Fig. 6b) and strength (Fig. 11b), is fully compatible with the approach of Zaiser and Sanfeld [44], who argued from similitude principles that the increase of flow stress due to external size (i.e., $\frac{\tau_{eff}}{\tau_{pin}}$ in our case) should, for pure metals, only depend on the product $L\sqrt{\rho_f}$ (i.e., our $R$ for pure metal). Our results demonstrate that this can be extended to alloys, using an appropriate definition of the dimensionless parameter $R = \frac{L}{l}$.

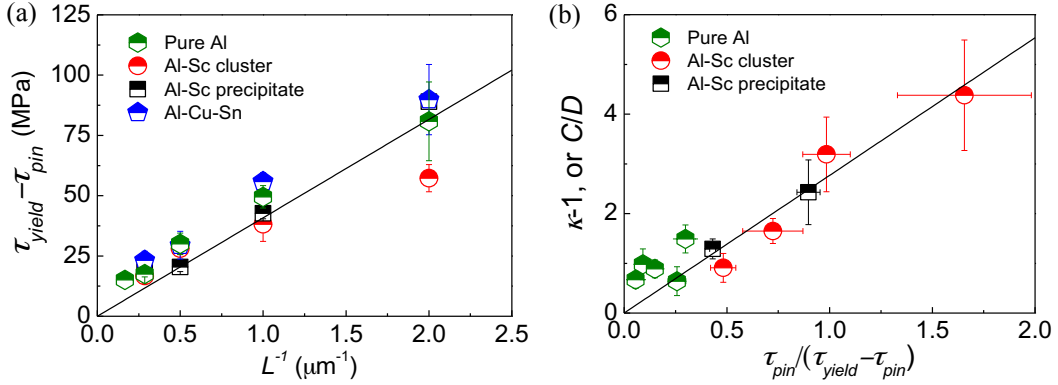

**Figure 11. The correlations between the power law exponent, $\kappa - 1 = C/D$, and size dependent effective shear stress, $\tau_{eff}$.** (a) The effective shear stress defined as $\tau_{eff} = \tau_{yield} - \tau_{pin}$, where $\tau_{yield}$ is the yield strength of micro-pillars estimated as the stress at 0.2% of plastic strain, as a function of external size $L$. The relation $\tau_{eff} \sim 1/L$ can be identified. (b) The power law exponent, $\kappa - 1$, as a function of $\tau_{pin}/(\tau_{yield} - \tau_{pin})$.

## 7. Concluding Comments

We performed compression tests on micro-pillars of pure Aluminum and Al-alloys single crystals strengthened by different types of solutes or precipitates, with diameters ranging from 500 to 6000 nm, and quantified the nature of plastic fluctuations occurring during deformation. From these data, we showed:

(1) Diminishing the external length scale (miniaturization) intensifies fluctuations and contributes to criticality. This is the "smaller is wilder" effect, that we quantified through the degree of wildness $W$ (the proportion of plastic deformation occurring through plastic avalanches) as well as through the time averaged scaling exponent of the avalanche size distributions.

(2) Introducing quenched disorder shifts the transition from "wild" to "mild" plasticity towards smaller external length scales. This illustrates the "dirtier is milder" effect, and opens the possibility to mitigate plastic instabilities at small length scales.

(3) The interplay of these effects reveal itself through a universal (material-independent) relation between the power law exponent of avalanche size distribution and the degree of wildness.

(4) By translating the pinning strength of obstacles into a characteristic length scale $l$, we showed that a single nondimensional parameter $R = L/l$ not only distinguishes the transition from source-exhaustion hardening to forest hardening, but also controls the wild to mild transition and determines the scaling exponent of avalanche size distribution.

(5) While jerky deformation and high flow stresses have been simultaneously observed at micro- and nano-scales by many authors, there has been so far no established link between them, either theoretically or experimentally. We now quantify this link through our dimensionless parameter $R$ which appears to be controlling both phenomena.

(6) A mean field model for the stochastic evolution of mobile dislocation density is proposed, which successfully recovers the universal relationship between the scaling exponent and the degree of wildness. It also rationalizes the competition between external (size related) and internal (disorder related) length scales on plastic fluctuations as well as the link between the size effects on strength and wildness.

In summary, our results provide a new unifying perspective on micro-plasticity linking together a broad range of relevant phenomena including wildness, avalanche size distribution, hardening mechanisms, strain heterogeneity, and high flow stresses.

**Acknowledgements**: Financial support for this work was provided by the Natural Science Foundation of China (Grant Nos. 51321003, 51322104, 51201123 and 51571157), the 111 Project of China (Grant No. B06025) and the French ANR-2008 grant EVOCRIT. J.W. and L. T. acknowledge the hospitality of the Aspen Center for Physics, supported by the NSF Grant
No. PHY-1066293.

# Supplement Materials (SM)

## 1. Statistics of wild fluctuations

We explicitly defined an "avalanche" as a plastic process characterized by a dissipation rate much greater than the imposed loading rate. Based on this definition and in view of the control mode provided by the loading device (Hysitron Ti 950), a dislocation avalanche manifests itself by both (i) a displacement/strain burst away from the displacement/strain-rate target (Fig. S1a, b and Fig. S2a) and (ii) a brutal stress drop (Fig. S1a, c and Fig. S2c). These two different manifestations can be separately used to distinguish wild from mild fluctuations, following a similar methodology (Fig. S2). The results obtained via the two methods are very similar (comparing Fig. 4 in the main text with Fig. S3).

The separation of wild from mild fluctuations is performed in two successive steps. The first step consists in extracting bursts (or avalanches) from a mild "noise". Once these bursts are extracted, a statistical analysis of the distributions of burst sizes $X$, based on a maximum likelihood methodology [S1], provides a lower bound $X_{min}$ to power law scaling for the burst size distribution. Only the bursts with size $X \geq X_{min}$ are considered as wild fluctuations. The cumulative effect of these wild fluctuations, normalized by the total imposed deformation, defines the wildness $W$.

(1) The response of testing machine during avalanches

The testing machine we used (Hysitron Ti 950) is inherently a load-controlled system. A feedback control loop is used to impose a constant displacement rate by actively changing the loading rate. This control mode, which is the same as the one used in previous studies of this type [S2, 3, 4], works well for mild fluctuations, as the dissipation rate is comparable to the loading rate, but fails, by nature, to maintain a steady displacement/loading rate when wild fluctuation occur, leading to an abrupt strain burst or brutal loading drop.

A segment of a stress-strain curve of an Al-Sc cluster micro-pillar with 1000 nm in diameter is shown on Fig. S1 as an example, where each point represents a data point sampled at 200 Hz. An abrupt plastic event, which manifests itself by a strain burst, lasting two time intervals, i.e. 0.01 s, is shown in Fig. S1a. As the result of the sudden release of elastic energy induced by a plastic event, the control system urgently reduces the loading (i.e. brutal stress drop in Fig. S2c) to trace back to the target position. However, it still takes ~0.53 s to reach the desired stress level (Fig. S1b and c), in sharp contrast with the plastic event duration of 0.01 s. The hysteresis in system loading response during an avalanche makes a constant displacement/strain rate impossible to maintain, explaining the existence of strain bursts such as shown in Fig. S1b and Fig. S2a. It is then reasonable to consider either the displacement/strain burst, or the brutal load drop as a signature of a dislocation avalanche. The size of the avalanche is quantified by the axial plastic displacement over the duration of the strain burst: $X = D_e - D_s + (F_s - F_e)/K_p$, where $D_s$, $F_s$, $D_e$, and $F_e$ are displacement and force when a burst starts and ends, respectively, as defined in Fig. S1, and $K_p$ is the stiffness of the sample. The axial displacement X during an avalanche is proportional to $b(\Delta A/A)$, where $\Delta A$ is the cumulative area swept by dislocations during the avalanche, A the slip area, and *b* the Burgers vector.

(2) Methodology of extracting wild avalanches

The following methodology has been adopted to extract displacement bursts and brutal load drops from the mild noise. A representative displacement rate vs time signal, where the target displacement rate has been removed, is shown in Fig. S2a. The peaks, which are responsible for the asymmetry of the fluctuations with respect to the horizontal axis, are signatures of strain bursts. The distribution of positive displacement rates, measured at the sampling frequency of 200 Hz, is compared with that of the negative ones (in absolute value) on Fig. S2b in order to determine the threshold where the two distributions diverge. For the positive displacement rates, a power-law tail is obtained. Below the threshold value, the two distributions coincide and are Gaussian-like, which means that all the asymmetry can be explained by the power-law tail of the positive part. This threshold can then be used to distinguish avalanches from background noise and slow dissipations.

Following an approach similar to that mentioned above for displacement-rate signals, we determined a threshold from the asymmetry of force/stress-rate distributions (Fig. S2c and d). The size of the avalanche is then estimated as the load drop (marked in Fig. S2c) divided by the pillar stiffness, which has the same physical meaning as the axial plastic displacement during an avalanche. The statistical results derived from the force/stress-based method, shown in Fig. S3, are in excellent agreement with the results derived from the displacement/strain-based method (see Fig. 4 in the main text).

(3) Correction for wild fluctuations.

Once these bursts are extracted from the mild noise, their distributions can be analyzed (see Fig. 5 of the main text). For some of these distributions, it is clear that power law scaling breaks down below a threshold $X_{min}$. This threshold can be determined by means of maximum likelihood methodology used to estimate the power law exponent $\kappa$ [S1]. Consequently, only the events with size X above $X_{min}$ are considered as true wild fluctuations, and taken into account in the estimation of wildness. However, it is worth noting that this second step induces only a higher order correction to the value of wildness and does not change any of the general trends.

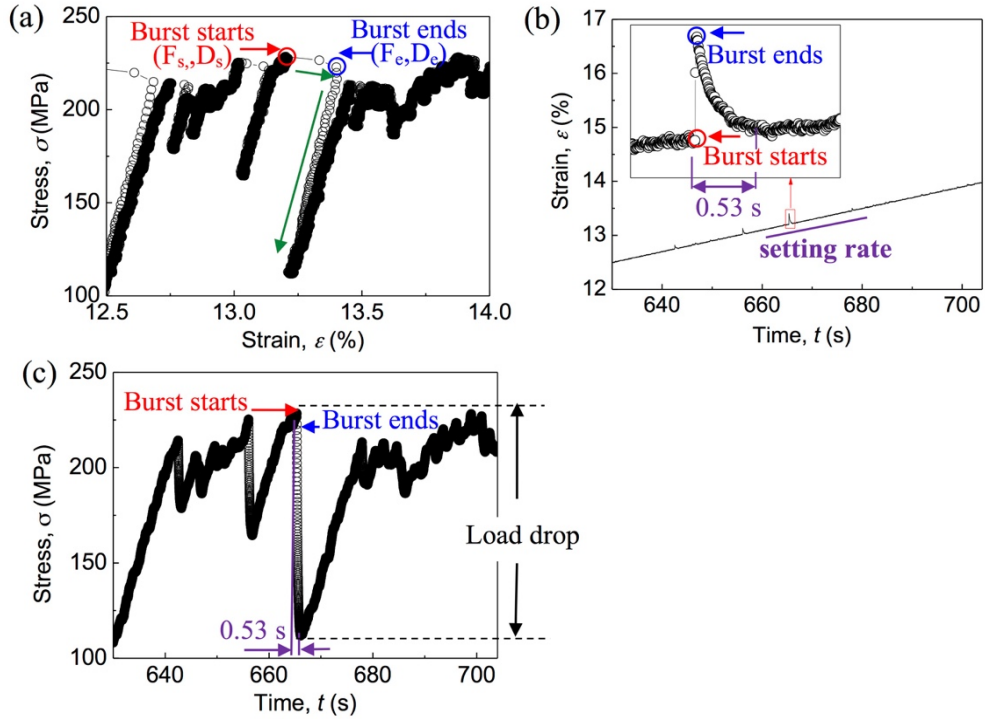

**Figure S1. Zoomed deformation curves of an Al-Sc cluster micro-pillar with 1000 nm diameter. (a)** A segment taken from the stress-strain curve, where a strain burst is marked by its starting (red circle) and ending point (blue circle). The sampling time is 0.005 s, which means that the strain burst is finished within 0.01 s, followed by an elastic unloading. The strain burst and the following elastic unloading are indicated by green arrows. **(b)** The corresponding strain-time curve. The strain burst marked in (a) is shown in a zoomed window to see the jump event and the retracting phase more clearly. **(c)** The corresponding stress-time curve. The burst starting and ending points are marked by red and blue arrows, respectively. It takes 0.53 s for the loading feedback loop to reach the target position. This means that the rate of wild fluctuation (finished in 0.01 s) is much faster than the loading reaction rate (taking 0.53 s). The hysteresis in system loading reaction leads to the strain burst in (a) and (b).

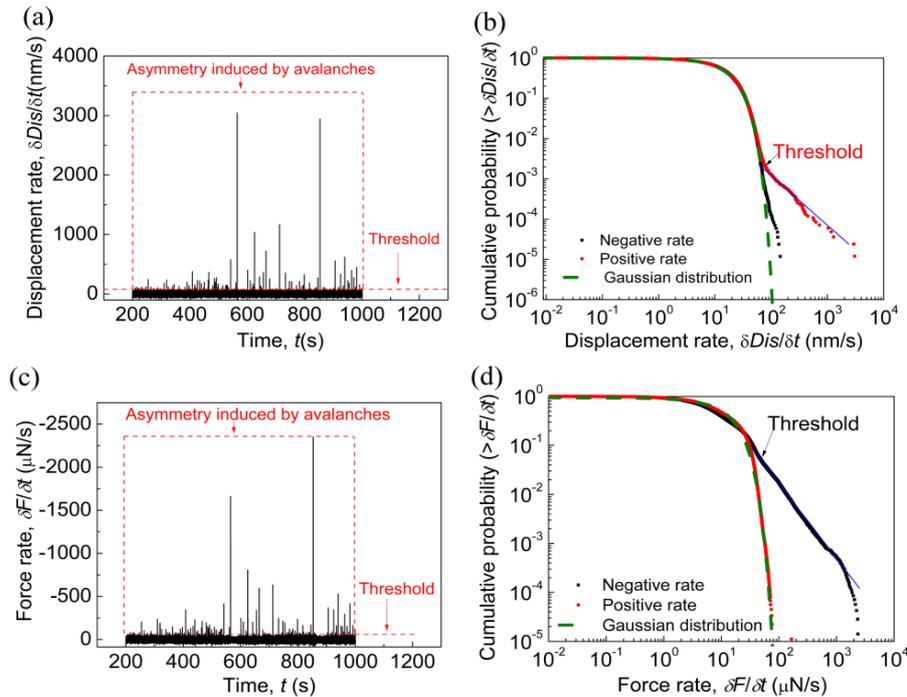

**Figure S2. Displacement/force-rate signals and corresponding distributions. (a)** Displacement rate-time signal

(after subtraction of the target rate) of an Al-Sc cluster micro-pillar with 1000 nm diameter. The asymmetry of the signal is a signature of wild fluctuations. **(b)** The cumulative probability of displacement positive and negative rates (in absolute value). The asymmetry of the signal is reflected by the power law tail of the positive rate distribution. Below the threshold, the two curves coincide, both following a Gaussian distribution. **(c)** and **(d)** are force rate-time signal and the cumulative probability of force positive and negative rates, respectively. In this case, the power law tail on the negative rates (stress drops) is the signature of wild fluctuations.

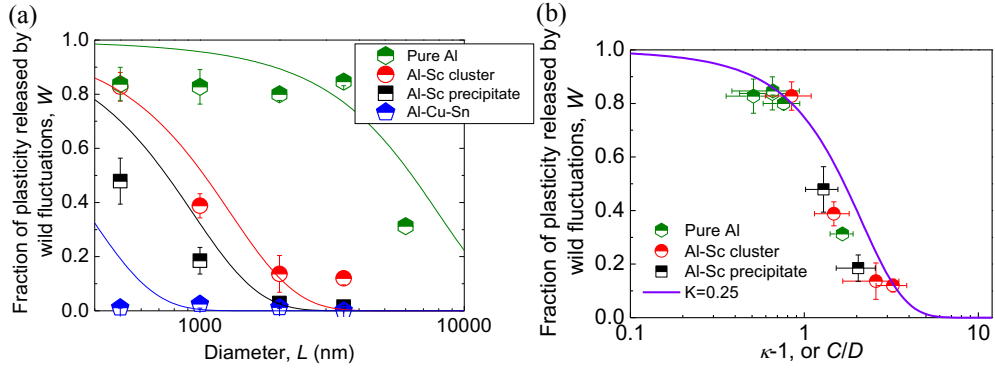

**Figure S3. The evaluation of $W$ and $C/D$ based on the force-rate signal. (a)** The external size effect on the degree of wildness $W$, and **(b)** the relationship between $W$ and the power law exponent $\kappa - 1$, or $C/D$. The curves in (a) and (b) are in close agreement with Fig. 4a and Fig. 4b ($K = 0.25$).

## 2. Simulations

To illustrate the signal generated by the model, we integrate the stochastic differential equation $\frac{d\rho_m}{d\gamma} = A - C\rho_m + \sqrt{2D}\,\rho_m \xi(\gamma)$, using Euler-Heun method [S5, 6] with a strain step $d\gamma$ of $10^{-6}$. Our choice of the numerical scheme corresponds to Stratonovich interpretation that we adopted to find the analytical solution in the steady regime. The parameters $C$ and $D$ are estimated for two extreme cases, namely 2000 nm-sized pure Al micro-pillar (wild) and 3500 nm-sized Al-Sc precipitate alloy (mild). From section 6.2.2 in the main text, we estimate $C \approx 400$ and $D \approx 800$, i.e. $C/D \approx 0.5$ for pure Al micro-pillar with 2000 nm diameter and $C \approx 700$ and $D \approx 100$, i.e. $C/D \approx 7$ for Al-Sc precipitate micro-pillar with 3500 nm diameter. To set the value of the last parameter, $A$, and to build the signals, we then need to set a value for the steady-state mobile dislocation density [S7] $\rho_{ss} = A/C$. For 3500 nm-sized Al-Sc precipitate micro-pillar, Taylor's hardening is at work (Fig. 8), hence the steady-state forest dislocation density $\rho_f$ can be roughly estimated from the flow stress at 5% strain, $\tau_{5\%} - \tau_{pin} = \alpha G b \sqrt{\rho_f}$, giving $\rho_f \approx 4\times 10^{13}$ m$^{-2}$. We therefore considered $\rho_{ss} \approx 10^{13}$ m$^{-2}$ as a reasonable estimation in this case, giving $A \approx 7\times 10^{15}$ m$^{-2}$. For pure Al, Ngan et al. [S8] measured a total dislocation density of $6.2\times 10^{12}$ m$^{-2}$ within a pillar of ~900 nm compressed along the same direction as our samples after 4% strain. Hence taking $\rho_{ss} \approx 10^{12}$ m$^{-2}$ in this case, it yields $A \approx 4\times 10^{14}$ m$^{-2}$. Note that these values have to be considered only as reasonable orders of magnitude, to give illustrative examples of the signals generated by our model.

The illustrations of mobile dislocation density evolution are shown in Fig. S4, where the signals transform from essentially Gaussian for Al-Sc precipitate ($C/D \approx 7$) to a strongly intermittent one for pure Al ($C/D \approx 0.5$), as in our experiments.

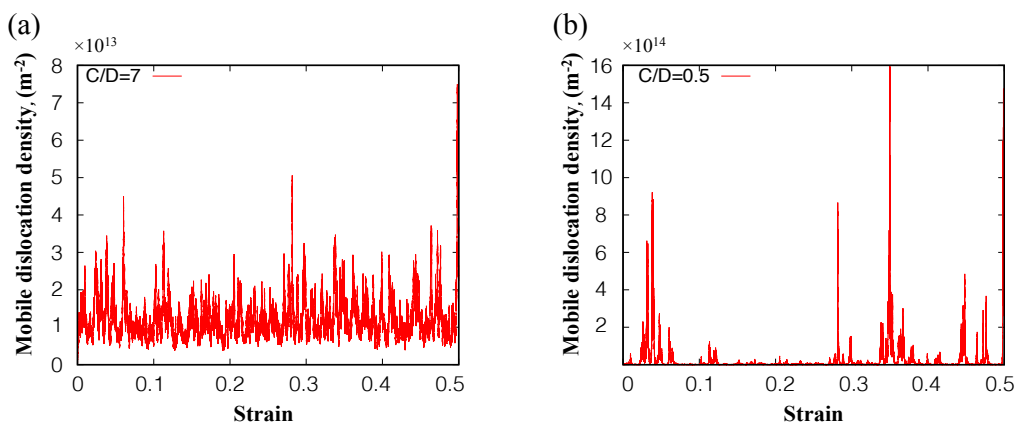

**Figure S4. Evolution of the mobile dislocation density throughout deformation simulated by our model**. Taking *C* and *D* values for *mild* 3500 nm-sized Al-Sc precipitate micro-pillar **(a)** and *wild* 2000 nm-sized pure Al micro-pillar **(b)** estimated from section 6.2.2 in the main text. The steady-state mobile dislocation densities are roughly estimated in section 6.4, and should be considered as reasonable orders of magnitude for simple illustrations only.